\documentclass[journal,english,twocolumn]{IEEEtran}

\usepackage[T1]{fontenc}
\usepackage{amsmath}
\usepackage{graphicx}
\usepackage{amssymb}
\usepackage{subfigure}
\usepackage{epstopdf}
\usepackage{gensymb}
\usepackage{cite}
\usepackage[dvipsnames]{xcolor}

\makeatletter
\usepackage[font=footnotesize]{caption}
\makeatother

\usepackage{tikz}
\newcommand\copyrighttext{%
  \footnotesize \textcopyright 2017 IEEE. Personal use of this material is permitted.
  Permission from IEEE must be obtained for all other uses, in any current or future 
  media, including reprinting/republishing this material for advertising or promotional 
  purposes, creating new collective works, for resale or redistribution to servers or 
  lists, or reuse of any copyrighted component of this work in other works. }
\newcommand\copyrightnotice{%
\begin{tikzpicture}[remember picture,overlay]
\node[anchor=south,yshift=0pt] at (current page.south) {\fbox{\parbox{\dimexpr\textwidth-\fboxsep-\fboxrule\relax}{\copyrighttext}}};
\end{tikzpicture}%
}

\usepackage{babel}

\usepackage[final]{changes} 

\colorlet{Changes@Color}{blue}
\begin{document}
\bstctlcite{IEEEexample:BSTcontrol}

\title{\added{Revisiting Hybrid} Interferometry with Low-Frequency Radio Astronomy Arrays}

\author{Adrian T. Sutinjo, Daniel Ung, Tim M. Colegate, Randall B. Wayth, Peter J. Hall, Eloy de~Lera Acedo 

\thanks{\textbf{IEEE Trans. Antennas. Propagat., accepted, 23 May 2017}}
}

\maketitle
\copyrightnotice

\begin{abstract}
Radio interferometry most commonly involves antennas or antenna arrays of identical design. The identical antenna assumption leads to a convenient and useful mathematical simplification resulting in a scalar problem. An interesting variant to this is a ``hybrid'' interferometer involving two designs. We encounter this in the characterization of low-frequency antenna/array prototypes using a homogenous low-frequency array telescope such as the Murchison Widefield Array (MWA). In this work, we present an interferometry equation that applies to hybrid antennas. The resulting equation involves vector inner products rather than scalar multiplications. We discuss physical interpretation and useful applications of this concept in the areas of sensitivity measurement and calibration of an antenna/array under test using a compact calibrator source.
\end{abstract}


\thispagestyle{empty}

\section{Introduction}
\label{sec:intro}

\added{M. Ryle in~\cite{Ryle351} provided a seminal treatment of radio interferometry with antennas of two different designs. This concept was later implemented in the Covington-Broten interferometer which consisted of two cylindrical parabolic dishes and a slotted waveguide antenna~\cite{1144510}, operated at 3~GHz. In low-frequency radio astronomy, an example of hybrid interferometry is the Mills~Cross Array~\cite{1953AuJPh...6..272M} comprised of a N-S and a E-W arrays of full-wave and folded dipole antennas operating at $\sim$97~MHz. The motivation for these early explorations in hybrid-antenna interferometry seems to be beam-shaping of the resultant interferometer response which was understood, at the time, as the scalar multiplication of the antenna voltage patterns and a fringe pattern that depends on the spacing between the antenna phase centers~\cite{1144510, 1448754}.}

\added{In recent years, radio astronomy interferometry has generally advanced from two (or few) elements to large number of elements. We most commonly find interferometers that consist of antennas with identical design, which simplifies the mathematics of imaging and calibration~\cite{1999ASPC..180.....T, tms01}. We refer to this as interferometry with ``homogeneous'' antennas. Examples of radio interferometry with ``heterogeneous'' elements, involving two or more antenna designs, are found in very-long-baseline interferometry (VLBA)~\cite{2009PASA...26...75P} and millimeter-wave astronomy~\cite{Wright_ALMA_2012}. At low-frequencies, an example of a somewhat heterogeneous interferometer is LOFAR high band (HBA)~\cite{2013A&A...556A...2V} which consists of stations (randomly) rotated relative to local north, but with bow-tie dipoles (of identical design) rotated back to be aligned with N-S. In our current context, ``hybrid'' interferometry involves antennas of two different designs which is useful for in-situ calibration and characterization of a prototype radio telescope. Recent examples include APERTIF phased array feed~\cite{6328718} and Low-Frequency Square Kilometre Array (SKA-Low) prototypes~\cite{7293140, 7303648}.}

\added{This paper revisits the theory of hybrid radio interferometry following the Jones matrix formalism~\cite{Smirnov:2011vp} and distills the implications to low-frequency antenna array calibration and characterization. Sec.~\ref{sec:review} demonstrates that the scalar multiplication result described in~\cite{1144510, 1448754} arises under certain conditions. We suggest a more general vector inner product form that lends itself readily to low-frequency array calibration where the antenna radiation patterns may be complex (i.e., involves amplitude and phase variations) and where the location of the phase center may not be evident. Physical interpretation of the vector form will be addressed in Sec.~\ref{sec:AUTcal}. Sec.~\ref{sec:example} illustrates successful application of the vector inner product form to astronomical calibration of a single antenna in a low-frequency array and to antenna/array under test (AUT) sensitivity measurement.} Concluding remarks are given in Sec.~\ref{sec:concl}.

\section{Review of Theory}
\label{sec:review}
We begin with a brief review of the ``measurement equation''~\cite{Smirnov:2011vp, HamakerI_1996} formalism in radio interferometry. This approach preserves the vector nature of the field quantities that is necessary to describe the hybrid array. For simplicity and insight, we prefer the $2\times2$ matrix approach~\cite{Smirnov:2011vp} as opposed to the $4\times4$~\cite{HamakerI_1996}. 

\begin{figure}[htb]
	\begin{center}
	\includegraphics[width=2in]{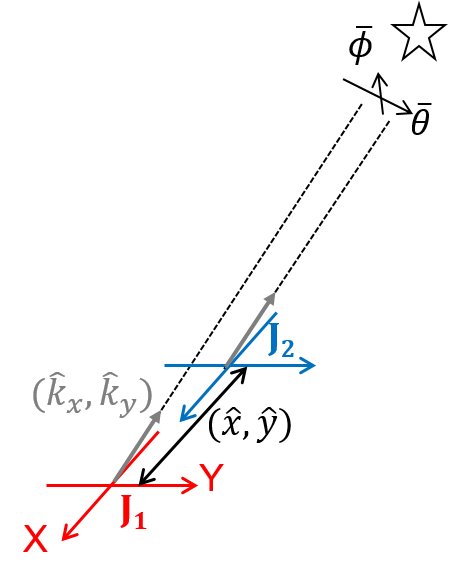}
	\end{center}
\caption{A hybrid two-``element'' interferometer. $\bar{\phi}$ and $\bar{\theta}$ are unit vectors in spherical coordinate system. $k_{x}=(2\pi/\lambda)\sin\theta \cos\phi$ and $k_{y}=(2\pi/\lambda)\sin\theta \sin\phi$ are the wavenumbers of the plane waves impinging on the interferometer and $(\phi,\theta)$ are angles in spherical coordinate system. The $\hat{.}$ refers to normalization to free-space wavenumber, $k_{0}=2\pi/\lambda$. $\mathbf{J}_{1}(\hat{k}_{x},\hat{k}_{y})$and $\mathbf{J}_{2}(\hat{k}_{x},\hat{k}_{y})$ are direction dependent Jones matrices for antennas $1$ and $2$, respectively. Note that we use ``antenna-centric'' coordinate system here as opposed to ``sky-centric'' one. $X$ and $Y$ refer to the orientation of antenna polarizations on the ground. $(\hat{x},\hat{y})$ refer to the location of antenna $2$ relative to antenna $1$, normalized to wavelength.}
\label{fig:1}
\end{figure}

A two-``element'' hybrid interferometer is depicted in Fig.~\ref{fig:1}. Note that each interferometer ``element'' could be a single antenna or an array of antennas whose voltages are combined. The Jones matrices of the arrays, $\mathbf{J}_{1}(\hat{k}_{x},\hat{k}_{y})$ and $\mathbf{J}_{2}(\hat{k}_{x},\hat{k}_{y})$, relate the electric field vector in the sky to the voltages measured by the $X$ and $Y$-directed array. For brevity, dependence on $(\hat{k}_{x},\hat{k}_{y})$ will be shown as $(\hat{k})$ henceforth. 
\begin{eqnarray}
\mathbf{v}(\hat{k})&=&\mathbf{J}(\hat{k})\mathbf{e}(\hat{k}) \nonumber \\
\left[ \begin{array}{c}
v_{X}(\hat{k}) \\
v_{Y}(\hat{k}) 
\end{array} \right]
&=&
\left[ \begin{array}{cc}
J_{X\theta}(\hat{k}) & J_{\added{X}\phi}(\hat{k}) \\
J_{\added{Y}\theta}(\hat{k}) & J_{Y\phi}(\hat{k}) 
\end{array} \right]
\left[ \begin{array}{c}
e_{\theta}(\hat{k}) \\
e_{\phi}(\hat{k}) 
\end{array} \right]
\label{eqn:J}
\end{eqnarray}
The entries of a Jones matrix has a unit of antenna \added{effective length} or 1/antenna factor~\cite{Milligan_2005_ch1}. In (\ref{eqn:J}), we use linear polarization bases for the sky and antennas. Note that for antennas that are stationary with respect to ground such as a crossed dipole, $\mathbf{J}(\hat{k})$ cannot be assumed to be a diagonal matrix in general~\cite{6420896, 6774441}. 

The measurand of interest is the partially polarized source that is quantified by the correlation of the sky electric fields ($\hat{k}$ dependence is implied but not shown below).
\begin{eqnarray}
\mathbf{B}= \left\langle \mathbf{e}\mathbf{e}^{H}\right\rangle
=\left[\begin{array}{cc}
\langle e_{\theta}e_{\theta}^{*} \rangle & \langle e_{\theta}e_{\phi}^{*} \rangle \\
\langle e_{\phi}e_{\theta}^{*} \rangle & \langle e_{\phi}e_{\phi}^{*} \rangle
\end{array} \right]
\end{eqnarray}
The above quantity is sensed by taking the correlation of the received antenna voltages.
\begin{eqnarray}
\mathbf{V}_{12}= \left\langle \mathbf{v}_{1}\mathbf{v}_{2}^{H}\right\rangle
=\left[\begin{array}{cc}
\langle v_{X1}v_{X2}^{*} \rangle & \langle v_{X1}v_{Y2}^{*} \rangle \\
\langle v_{X2}v_{Y1}^{*} \rangle & \langle v_{Y1}v_{Y2}^{*} \rangle
\end{array} \right]
\end{eqnarray} 

Assuming that the plane-wave components for distinct $\hat{k}$ are uncorrelated and the interferometers are coplanar ($z=0$), the correlation of the antenna voltages is given by
\begin{eqnarray}
\mathbf{V}_{12}(\hat{x},\hat{y})\!=\!
\int_{\hat{k}}\mathbf{G}_{1}\mathbf{J}_{1}(\hat{k})\mathbf{B}(\hat{k})\mathbf{J}_{2}^{H}(\hat{k})\mathbf{G}_{2}^{*}e^{-j 2\pi (\hat{k}_{x}\hat{x}+\hat{k}_{y}\hat{y})} d\hat{k}
\label{eqn:visibility}
\end{eqnarray}
where limits of the integral are $\pm\infty$ (however, $\mathbf{B}(\hat{k})$ is non-zero \added{only in the upper hemisphere, $0\leq(\hat{k}_{x}^{2}+\hat{k}_{y}^{2})^{1/2}\leq1$})\added{;} $(\hat{x},\hat{y})$ are normalized to $\lambda$. Assuming electronic crosstalk between $X$ and $Y$ channels is properly suppressed, $\mathbf{G}$ is a diagonal matrix containing \emph{direction independent} complex electronic voltage gains associated with the $X$ and $Y$ antennas.
\begin{eqnarray}
\mathbf{G}
=\left[\begin{array}{cc}
g_{X}& 0 \\
0 & g_{Y}
\end{array} \right]
\end{eqnarray}

In practice, neither $\mathbf{G}$ nor $\mathbf{J}(\hat{k})$ is known and must be estimated and/or inferred from certain measurements. Subsequently, these quantities are removed from the measurement, leaving just the information from the sky. This is referred to as ``calibration.''  

To calibrate the interferometer, we assume that the instrument sees a single unpolarized, bright, and point-like source in the sky. An example in the southern sky is Hydra A (HydA) which is bright ($\sim300$ Jansky [Jy] at 160~MHz)\added{, unpolarized,}\footnote{\added{HydA is known to have extremely high (in the $\sim10^{3}$ to $\sim10^{4}$ radians/m$^{2}$) rotation measures (RM)~\cite{1993ApJ...416..554T} which means that it is depolarized in our observation frequency and bandwidth as per $\Delta\psi_{1}-\Delta\psi_{2}=(\lambda_{1}^{2}-\lambda_{2}^{2})\mathrm{RM}$~\cite{Tools_ch3}. For observation center frequency of 150~MHz and 40~kHz bandwidth, $\lambda_{1}^{2}-\lambda_{2}^{2}\approx3.2\times10^{-3}~\mathrm{m}^2$, such that $\Delta\psi_{1}-\Delta\psi_{2}\geq2\pi$ for RM$\geq2\times10^{3}$ radians/m$^{2}$. The combination of high intrinsic RM and the relatively low angular resolution of the MWA to discern regions in HydA with coherent polarized radiation means that we do not expect any significant polarized signal from this source.}} and may be considered point-like  for interferometer separations less than a few hundred wavelengths~\cite{7293140}. In addition, the interferometer separations should be sufficiently large ($\approx>30\lambda$)~\cite{7293140, Colegate_CAMA2014} such that the galactic noise, which is dominant at low frequencies ($\sim<250$~MHz), is uncorrelated. Assuming these conditions are met, (\ref{eqn:visibility}) simplifies to 
\begin{eqnarray}
\mathbf{V}_{12}(\hat{x},\hat{y})=
\mathbf{G}_{1}\mathbf{J}_{1}(\hat{k})\mathbf{J}_{2}^{H}(\hat{k})\mathbf{G}_{2}^{*}\frac{I}{2}e^{-j 2\pi (\hat{k}_{x}\hat{x}+\hat{k}_{y}\hat{y})} 
\label{eqn:calibration}
\end{eqnarray}
For an unpolarized source, $\mathbf{B}=I\mathbf{I}/2$ where $I$ is Stokes $I$ for the source~\cite{Wilson:52076, 1999ASPC..180..111C, SutOSu15} and $\mathbf{I}$ is the identity matrix. We further assume that a phase difference commensurate to the position of the source is applied to the interferometer such that the exponential terms cancel. 
\begin{eqnarray}
\mathbf{V}_{12}&=&
\mathbf{G}_{1}\mathbf{J}_{1}(\hat{k})\mathbf{J}_{2}^{H}(\hat{k})\mathbf{G}_{2}^{*}\frac{I}{2} \nonumber \\
&=&
\left[ \begin{array}{c c}
M(1,1)& M(1,2)\\
M(2,1) & M(2,2) \end{array} \right] \frac{I}{2}
\label{eqn:calibration2}
\end{eqnarray}
where the entries of the $\mathbf{M}$ are
\begin{eqnarray}
M(1,1)&=& g_{X1} (J_{X1\theta}J_{X2\theta}^{*}+J_{X1\phi}J_{X2\phi}^{*}) g_{X2}^{*} \nonumber \\
M(1,2)&=& g_{X1} (J_{X1\theta}J_{Y2\theta}^{*}+J_{X1\phi}J_{Y2\phi}^{*}) g_{Y2}^{*} \nonumber \\
M(2,1)&=& g_{Y1} (J_{Y1\theta}J_{X2\theta}^{*}+J_{Y1\phi}J_{X2\phi}^{*}) g_{X2}^{*} \nonumber \\
M(2,2)&=&g_{Y1} (J_{Y1\theta}J_{Y2\theta}^{*}+J_{Y1\phi}J_{Y2\phi}^{*}) g_{Y2}^{*} 
\label{eqn:calibration_M}
\end{eqnarray}
Although not explicitly shown in the matrix, recall that $g$ terms are direction independent while the $J$ terms are direction dependent.

\subsection{Identical Arrays}
\label{sec:identical}
The most convenient and standard form of the calibration equation occurs when the antennas/arrays are identical such that $\mathbf{J}_{1}(\hat{k})=\mathbf{J}_{2}(\hat{k})$. In this case, the diagonal entries of matrix $\mathbf{M}$ in (\ref{eqn:calibration2}) simplify to:
\begin{eqnarray}
M(1,1)&=&g_{X1} (|J_{X\theta}|^{2}+|J_{X\phi}|^{2}) g_{X2}^{*}\nonumber \\
M(2,2)&=&g_{Y1} (|J_{Y\theta}|^{2}+|J_{Y\phi}|^{2}) g_{Y2}^{*} 
\label{eqn:calibration_identical}
\end{eqnarray}
where $|J_{X\theta}|^{2}+|J_{X\phi}|^{2}=J_{X}^{2}$ is the square of the total antenna height for the $X$-directed antenna (similarly for the $Y$-directed antenna). 

\added{It is common practice}\footnote{\added{in standard radio astronomy software packages (e.g., https://casa.nrao.edu/). This remains a sensible approach in low-frequency interferometry calibration as the diagonal entries of $\mathbf{M}$ vary much more smoothly over $\phi$ and is higher near zenith (which coincides with the expected maximum array response) than the cross diagonals. Take the $\mathbf{J}_{1,2}$ as Hertzian dipoles~\cite{SutOSu15}, for example. Compare $M(1,1)\sim \cos^{2}\theta\cos^{2}\phi+\sin^{2}\phi$ (smooth with respect to $\phi$ and close to 1 near zenith) to $M(1,2)\sim-0.5\sin(2\phi)\sin^{2}\theta$ (undulates with $\phi$ and is close to zero near zenith).}} \added{in radio astronomy calibration} to treat $X$ and $Y$ antennas separately~\cite{1989ASPC....6...83F}, taking $gJ(\hat{k})$ as a single complex unknown with direction dependent amplitude but direction \emph{independent} phase \added{(consistent with the view that the antenna radiation pattern is \emph{real} and any phase term is purely a conducted phenomenon due to electronics and/or cable delays)}. This leads to the standard scalar equation (the subscript $_X$ or $_Y$ are suppressed from this point)
\begin{eqnarray}
\langle v_{1}v_{2}^{*}\rangle=g_{1}J(\hat{k}) J(\hat{k}) g_{2}^{*} I/2
\label{eqn:calibration_scalar}
\end{eqnarray}
With $N$ antennas/arrays, we obtain $N(N-1)/2$ equations with $N$ unknowns. Assuming the brightness of the calibrator source ($I$) is known, we get an overdetermined problem which may be solved using a linear least-squares method~\cite{1989ASPC....6..185C, 7293140}. 

\subsection{Hybrid Arrays}
\label{sec:hybrid}
For a hybrid array, $\mathbf{J}_{1}(\hat{k})\neq\mathbf{J}_{2}(\hat{k})$. Hence, $M(1,1)$ and $M(2,2)$ in (\ref{eqn:calibration_M}) do not simplify. Note that, for a hybrid array,
\begin{eqnarray}
\langle v_{1}v_{2}^{*}\rangle&\neq&g_{1} J_{1}(\hat{k})J^{*}_{2}(\hat{k})g_{2}^{*}I/2 
\label{eqn:calibration_identical3}
\end{eqnarray}
in general. The correct expression is
\begin{eqnarray}
\langle v_{1}v_{2}^{*}\rangle=g_{1} \mathbf{j}_{1}^{T}(\hat{k})\mathbf{j}_{2}^{*}(\hat{k})g_{2}^{*} I/2 
\label{eqn:calibration_non_vec}
\end{eqnarray}
where $\mathbf{j}_{1}^{T}=[J_{1\theta},~J_{1\phi} ]$ (similarly for $\mathbf{j}_{2}$); also, note that $g$ is a scalar scaling factor to $\mathbf{j}$. Hence, the vector inner product only simplifies to a scalar multiplication if the vectors are \emph{co-linear}, the physical meaning of which will be clarified in the next section. 

\section{AUT Calibration}
\label{sec:AUTcal}
In (\ref{eqn:calibration_non_vec}), let the subscript $_{1}$ denote the AUT. This equation may be re-written as~:
 \begin{eqnarray}
\langle v_{1}v_{2}^{*}\rangle&=& |\langle v_{1}v_{2}^{*}\rangle|e^{j\angle\langle v_{1}v_{2}^{*}\rangle}\nonumber \\
&=&g_{2}^{*}\left\|\mathbf{j}_{2}\right\| \left\| \mathbf{j}_{1}\right\| g_{1}\cos\alpha~e^{j\gamma} I/2 
\label{eqn:AAVS_MWA_3}
\end{eqnarray}
where $g||\mathbf{j}||=g\sqrt{\mathbf{j}^{H}\mathbf{j}}$, $\alpha$ is the angle between the two complex vectors, $\angle\langle v_{1}v_{2}^{*}\rangle=\angle(\mathbf{j}_{2}^{H}\mathbf{j}_{1})+\angle (g_{2}^{*}g_{1})$ and $\gamma=\angle(\mathbf{j}_{2}^{H}\mathbf{j}_{1})$.

Given the form of (\ref{eqn:AAVS_MWA_3}) and assuming that $\mathbf{j}_{2}$ to $\mathbf{j}_{\added{N}}$ are identical antennas, the linear least square procedure mentioned in Sec.~\ref{sec:identical} may be applied. However, the solution obtained for the AUT will be $g_{1} \left\|\mathbf{j}_{1}\right\| \cos\alpha~e^{j\gamma}$ as opposed to $g_{1}\left\|\mathbf{j}_{1}\right\|$. This is demonstrated below with $N=3$ for clarity (note $\mathbf{j}_{2} =\mathbf{j}_{3}$).
\begin{eqnarray}
\langle v_{1}v_{2}^{*}\rangle&=&g_{2}^{*}\left\|\mathbf{j}_{2}\right\| \left\| \mathbf{j}_{1}\right\| g_{1}\cos\alpha~e^{j\gamma} I/2  \nonumber \\
\langle v_{1}v_{3}^{*}\rangle&=&g_{3}^{*}\left\|\mathbf{j}_{2}\right\| \left\| \mathbf{j}_{1}\right\| g_{1}\cos\alpha~e^{j\gamma} I/2 \nonumber \\
\langle v_{2}v_{3}^{*}\rangle&=&g_{3}^{*}\left\|\mathbf{j}_{2}\right\|\left\|\mathbf{j}_{2}\right\|g_{2} I/2
\label{eqn:AAVS_MWA_linear_eq}
\end{eqnarray}
There are 3 complex unknowns in (\ref{eqn:AAVS_MWA_linear_eq}): $g_{2}\left\|\mathbf{j}_{2}\right\|$, $g_{3}\left\|\mathbf{j}_{2}\right\|$, and $ g_{1} \left\| \mathbf{j}_{1}\right\|\cos\alpha~e^{j\gamma}$. 

Note the physical meanings of the following parameters: 
\subsubsection{$\cos \alpha$}
\begin{figure}[htb]
	\begin{center}
	\includegraphics[angle =270, width=3.7in]{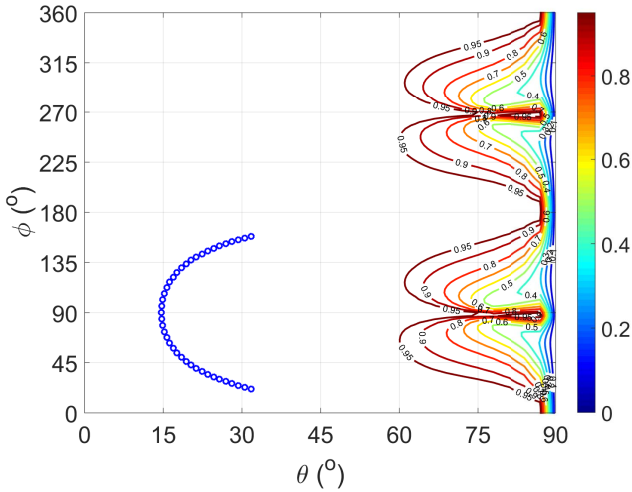}
	\end{center}
\caption{FEKO simulated $\cos\alpha$ factor for a single SKALA and a single MWA bow-tie at 220 MHz. The AUTs are dual-polarized linearly polarized elements oriented along N-S (``Y'') and E-W (``X''). The results above were obtained by exciting the ``Y'' element while keeping the ``X'' element open circuited. The SKALA antenna is placed on a soil model with 2\% moisture~\cite{7293140} and the MWA bow-tie is placed over a perfect electric ground plane. The coordinate origins are placed such that the feed is located at (0,0,$z$). The $(\phi,\theta)$ in the plots are that of the spherical coordinate system. The blue \added{circles} represent the trajectory of HydA down to zenith angle ZA$\sim30^{\circ}$. \added{The contour lines are 0.95, 0.9, 0.8, ...0.1.}}
\label{fig:cos_220}
\end{figure}

\begin{eqnarray}
\cos\alpha=\frac{|\mathbf{j}_{2}^{H}\mathbf{j}_{1}|}{\left\|\mathbf{j}_{2}\right\| \left\|\mathbf{j}_{1}\right\| }
\label{eqn:cos_alp}
\end{eqnarray}
$(\cos\alpha)^2$ is a direction dependent polarization mismatch factor~\cite{Stutz_2013_ch4} of the AUT and the identical antennas/arrays. Fig.~\ref{fig:cos_220} illustrates the simulated $\cos \alpha$ factor obtained by taking the normalized complex inner product in (\ref{eqn:cos_alp}) of a single log-periodic antenna prototype for SKA-Low (SKALA)~\cite{Eloy_EXAP2015} and a single MWA bow-tie antenna~\cite{Lonsdale_2009,2013PASA...30....7T} at 220~MHz. The blue dots illustrate a portion of the HydA's trajectory down to $\sim30^{\circ}$ zenith angle. At this frequency, polarization mismatch is small ($\cos\alpha>0.95$) for $\theta<60^{\circ}$. 

\subsubsection{$\gamma$}
\begin{eqnarray}
\gamma=\angle(\mathbf{j}_{2}^{H}\mathbf{j}_{1})
\label{eqn:gamma}
\end{eqnarray}
is a direction dependent phase term due to the displacement of the phase-center of the AUT relative to that of the identical antennas/arrays beyond the $e^{-j 2\pi (\hat{k}_{x}x+\hat{k}_{y}y)}$ interferometer separation factor in (\ref{eqn:calibration}). 

\begin{figure}[htb]
	\begin{center}
	\includegraphics[angle =270, width=3.7in]{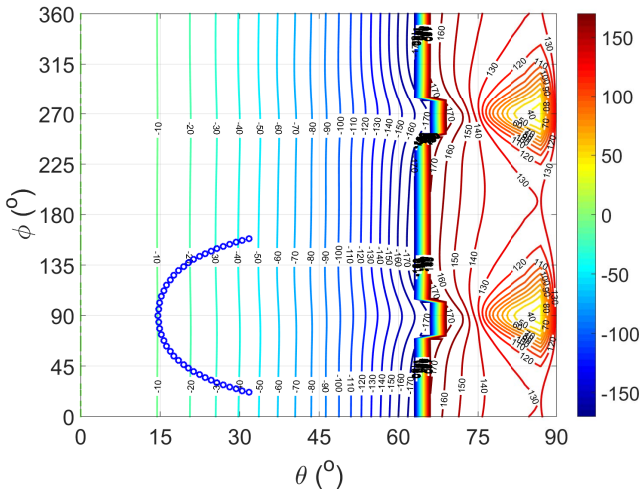}
	\end{center}
\caption{FEKO simulated $\gamma$ factor (degrees) for a single SKALA and a single MWA bow-tie at 220 MHz. Simulation parameters are identical to Fig.~\ref{fig:cos_220}. The phase at $\theta=0^{\circ}$ is set to zero. The contour step size is $10^{\circ}$.}
\label{fig:gamma_220}
\end{figure}

Simulated $\gamma$ factor of a single SKALA and an MWA bow-tie is shown in Fig.~\ref{fig:gamma_220}. Note that over the trajectory of HydA indicated on the figure, the $\gamma$ factor changes by approximately $35^{\circ}$. For $\theta<60^{\circ}$, the $\gamma$ factor is nearly azimuthally symmetric with approximately linear slope as a function of $\cos\theta$ (for $\theta\lesssim50^{\circ}$). As expected of a log-periodic antenna, this behavior is indicative of the height of the SKALA's phase center relative to the MWA bow-tie. 

The phase-center offset may be inferred by taking the derivative of $k_{z}\Delta z=(2\pi/\lambda)\Delta z \cos\theta$ with respect to $\theta$ and solving for $\Delta z$~\cite{1305530}. Following this method, we obtain $\Delta z \approx 1.04$~m for Fig.~\ref{fig:gamma_220}. At 220~MHz, the physical height of the half-wavelength element above the base of the SKALA is $\Delta z \approx 1.1$~m, consistent with calculation. 


In summary, assuming a single bright compact calibrator source, complex gain calibration in hybrid interferometry (1 AUT + $N-1$ identical antenna/arrays) results in direction-dependent factors in \emph{both} amplitude and phase:
\begin{itemize}
\item The amplitude factor is due to polarization mismatch of the AUT and the identical antenna/array. This may be minimized by polarization matching of the radiated far-fields. Conversely, if the polarization mismatch is known from electromagnetic (EM) analysis/simulation, this effect may be corrected.
\item The phase factor is due to relative movement of AUT's phase center with respect to the $N-1$ identical antenna/array. This factor could be reduced by minimizing the relative phase-center offset. However, as AUT design do not typically consider this factor, this option is likely impracticable. Hence, we rely on EM analysis of the AUT and the identical antenna/array to correct this artifact. 
\end{itemize}
\added{Note that by writing the equation in the form  of \eqref{eqn:AAVS_MWA_3}, the location of the phase center need not be explicitly known. This is convenient for working with radiation patterns that are complex and/or known only through numerical simulations and where the phase center may vary as function of frequency, pointing angle, direction, and embedded position in an array. These aspects will be demonstrated} with AUT calibration \added{using} HydA in the next section. In addition, we discuss the implication of these findings on AUT sensitivity estimation.

\section{Example Applications}
\label{sec:example}

\subsection{Calibration of a Single AUT Using the MWA}
\label{sec:cal_single}
SKA-Low envisages using RF-over-Fiber (RFoF)~\cite{SKA1baseV2} to transport radio frequency (RF) signal with low loss over a few kilometers from each antenna to the digitizer. A digital beamformer takes these inputs and coherently sums them to form a phased array beam. This requires calibration of RFoF phase delay from each input. A coaxial based system such as the MWA~\cite{2013PASA...30....7T} relies on phase-matched cables connected to the inputs of its analog beamformer. However, fiber optic phase matching is not a commercially available option. \added{In this section, we explore hybrid interferometry as a plausible means to this calibration task. }    

We offer two examples based on observation of HydA as  calibrator source at the \added{Murchison Radio-astronomy Observatory (MRO)}. The first is phase calibration of a SKALA antenna embedded in a 16-element pseudo-random array (called AAVS0.5)~\cite{7293140} using the MWA $4\times4$ arrays (referred to as ``MWA tiles''). As a comparison, the second example is phase calibration of a single MWA bow-tie embedded in the $4\times4$ array with the MWA tiles. Both cases involve hybrid interferometry. Though less obvious, the latter is a hybrid problem because the radiation of an embedded bow-tie element is influenced by mutual coupling.     

\begin{figure}[tb]
	\centering
	\subfigure[]{
		\includegraphics[width=0.9\linewidth]{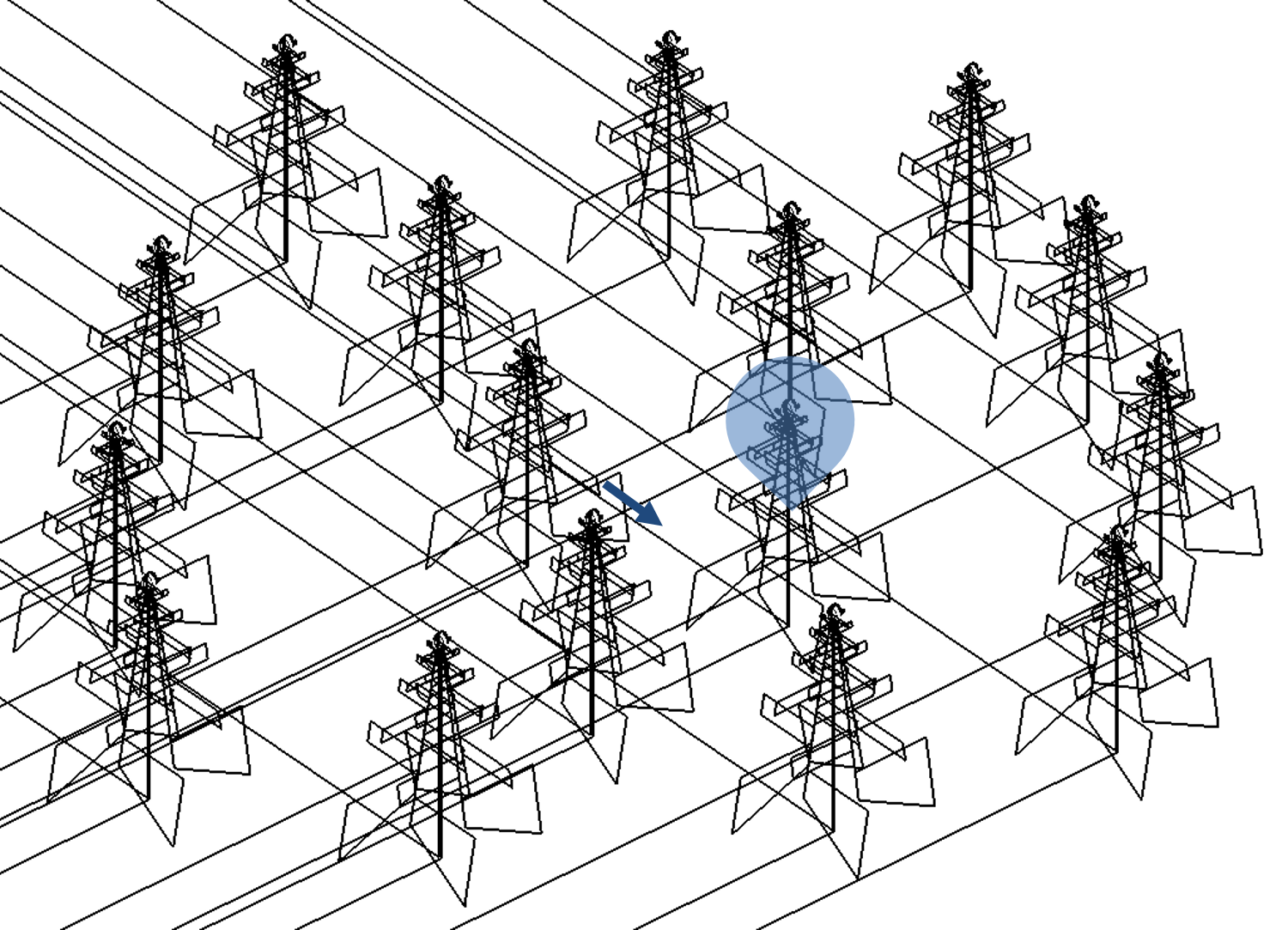}\label{fig:SKALA_embed11}} 
	\subfigure[]{
		\includegraphics[width=0.9\linewidth]{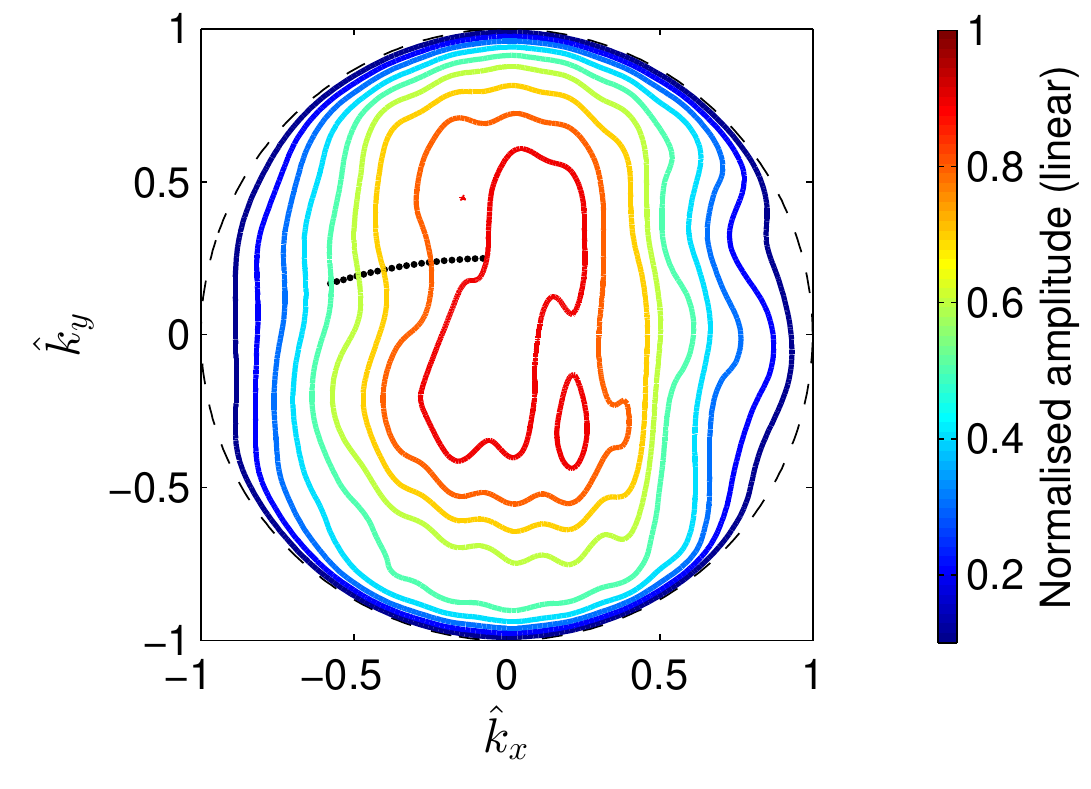}\label{fig:AAVS 2D beam}} 
	\caption{Embedded SKALA element location and beam pattern. ~\subref{fig:SKALA_embed11} The location of the embedded SKALA (referred to as number 11, indicated with a blue bubble) in the 16-element array. The SKALA 11 is located at 0.27~m to the west and 1.20~m to the north of the reference center of the array approximately denoted by the tail of the blue arrow. The head of the blue arrow points north.~\subref{fig:AAVS 2D beam} The normalized SKALA embedded element power pattern at 110~MHz. The contour step size is 0.1. Black dotted trace shows the observed HydA trajectory. For more detail regarding AAVS0.5/SKALA and MWA full-wave simulations please see~\cite{7293140} and~\cite{SutOSu15}, respectively  } 
	\label{fig:SKALA beams}
\end{figure}

\begin{figure}[htb]
	\centering
	\subfigure[]{
		\label{fig:AAVS cal soln}
		\includegraphics[width=0.9\linewidth]{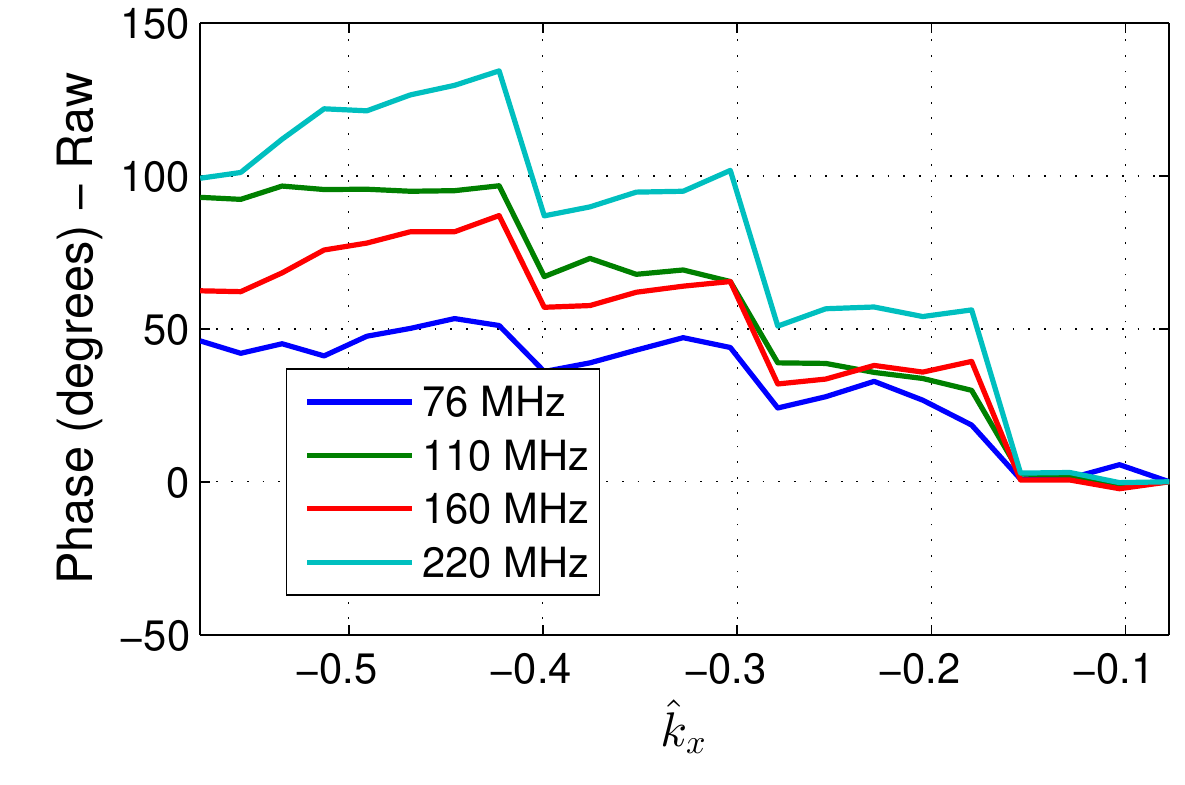}
		} 
	\subfigure[]{
		\label{fig:AAVS SMC}
		\includegraphics[width=0.9\linewidth]{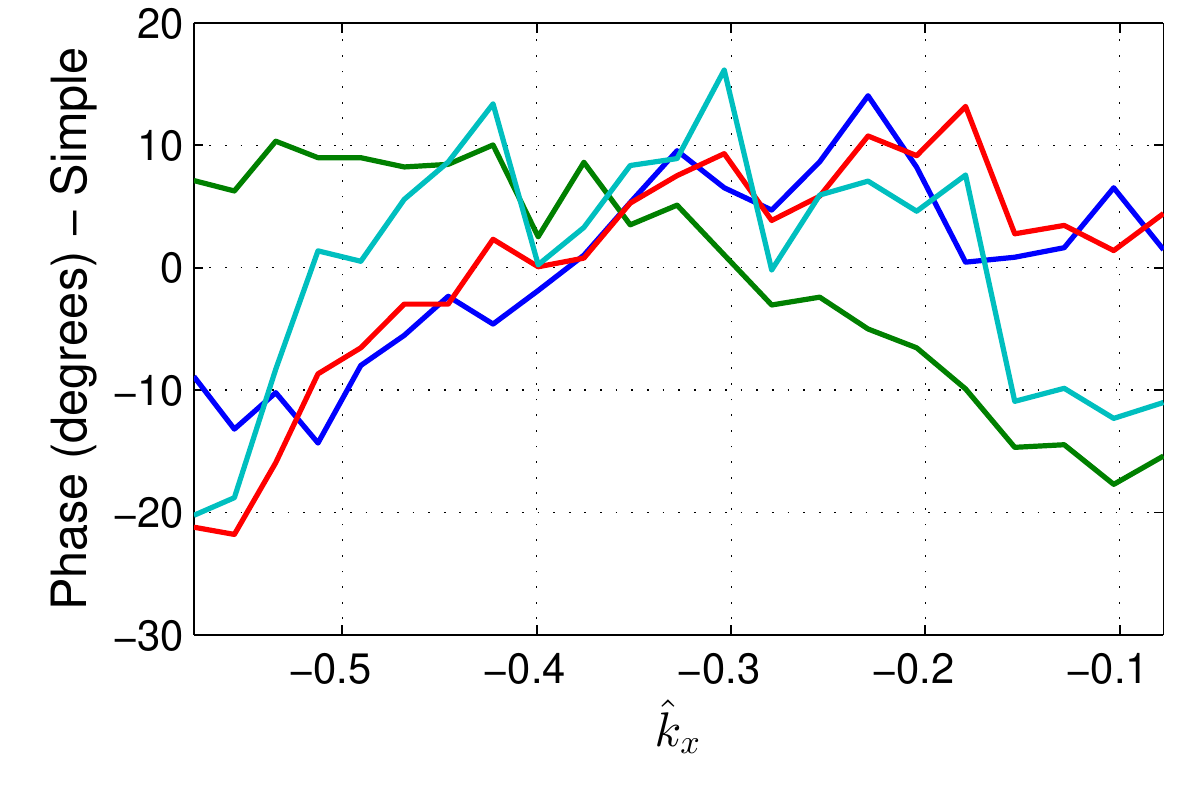}
		}
		
		\subfigure[]{
			\label{fig:AAVS GMC}
			\includegraphics[width=0.9\linewidth]{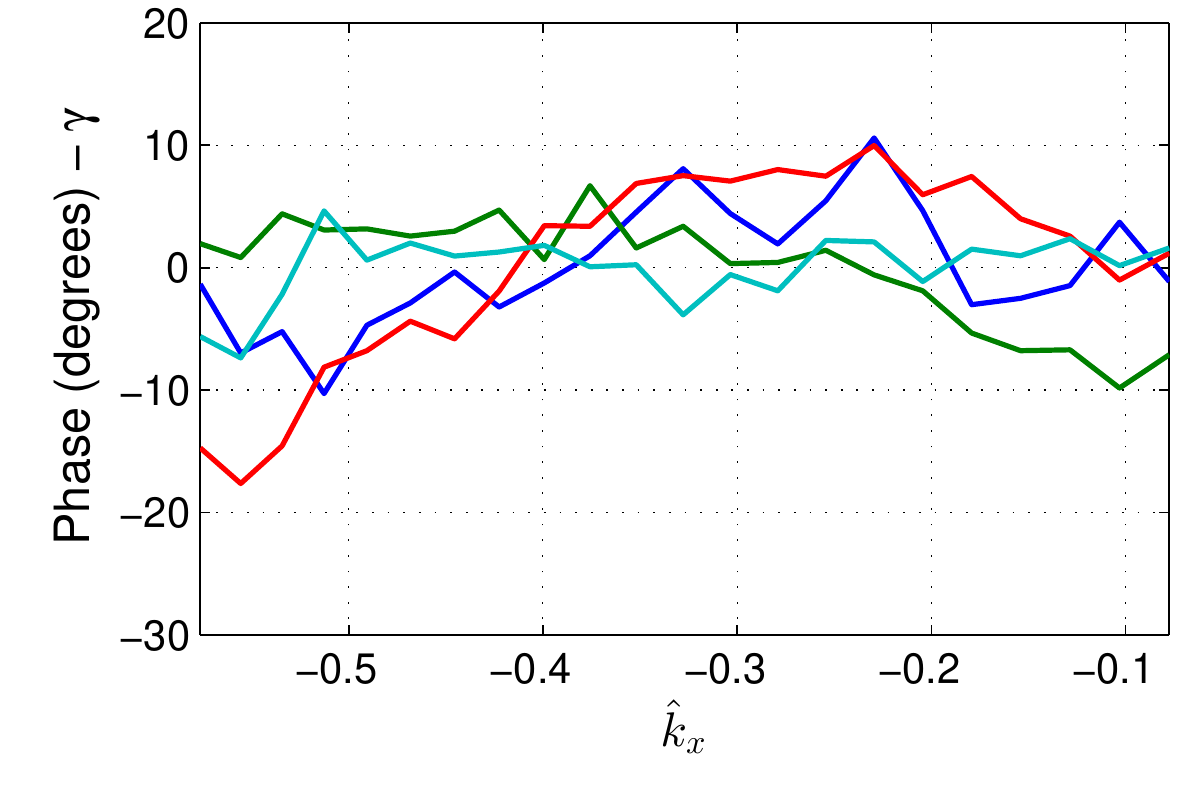}
		}
	
	\caption{Phase of the calibration solutions of SKALA~11-MWA tile hybrid interferometer with E-W orientation:~\subref{fig:AAVS cal soln} raw calibration solution,~\subref{fig:AAVS SMC} corrected phase solution using simple geometric model, ~\subref{fig:AAVS GMC} corrected using $\gamma$ factor. \label{fig:AAVS soln comparison}}
\end{figure}

Fig.~\ref{fig:SKALA beams} shows the location in the array and the  normalized embedded power pattern of the SKALA antenna (number 11) to be calibrated. The calibration strategy is to correlate the voltage of SKALA~11 with that of the MWA tile that tracks HydA along its trajectory. Fig.~\ref{fig:AAVS cal soln} reports the phase of the ``raw'' (uncorrected) calibration solution obtained via least squares (see discussion in Sec.~\ref{sec:AUTcal} regarding \eqref{eqn:AAVS_MWA_linear_eq}). \added{Recall from Sec.~\ref{sec:AUTcal} that the desired calibration solution should be  free from the direction dependent phase term $\gamma$. Hence, } several features in that figure merit further explanation. The most obvious are the phase steps corresponding to the switching of delay settings in the analog MWA beamformer as it re-points to track HydA. \added{These} values are known from instrument design~\cite{2013PASA...30....7T} and are easily corrected.

Less obvious features are the slight phase slopes at every pointing. This is contributed to by the geometric offset of SKALA~11 from the reference center of the array and the phase variation within the hybrid antenna beams. The geometric offset is known (see Fig.~\ref{fig:SKALA beams} caption) and also easily corrected by entering the location of the SKALA~11 in \eqref{eqn:calibration}. The result of this correction is shown in Fig.~\ref{fig:AAVS SMC}. Without knowledge of hybrid interferometry and full-wave EM simulation, this reflects the best calibration effort. In Fig.~\ref{fig:AAVS SMC}, we notice residual phase drift of up to $\sim30^{\circ}$. At 220~MHz, we observe remaining phase steps of $\sim15^{\circ}$ which \added{is consistent with} mutual coupling in the MWA tile~\cite{SutOSu15}.

Sec.~\ref{sec:AUTcal},  in particular \eqref{eqn:gamma}, suggests correction using the phase term of the inner product. We obtain this information from FEKO simulations of electric far-field vectors of SKALA~11 and the MWA tile along the trajectory of HydA. Fig.~\ref{fig:AAVS GMC} reports the phase after $\gamma$ term correction where we see improvement in the residual phase compared to Fig.~\ref{fig:AAVS SMC}. Note also that the residual phase steps at 220~MHz have been removed. Tab.~\ref{tab:AAVS std dev} summarizes the phase standard deviation of the calibration solution when corrected with a simple geometric model and $\gamma$ factors. There is generally a reduction\footnote{the only slight improvement using $\gamma$ correction at 160~MHz is noted. \added{Please see more discussions on this in the appendix.}} in phase spread when the $\gamma$ correction is applied.

\begin{table}[htb]
	\centering
	\caption{SKALA~11 and MWA tile hybrid calibration: standard deviation of phase residues.}
\begin{tabular}{|c|c|c|}
	\hline
	Frequency & Geometric & $\gamma$ correction \\
	\hline
	75.5 MHz &$7.75^{\circ}$ & $5.04^{\circ}$ \\
	\hline
	110 MHz &$9.45^{\circ}$ & $4.44^{\circ}$ \\
	\hline
	160 MHz &$9.65^{\circ}$ & $8.23^{\circ}$ \\
	\hline
	220 MHz &$10.22^{\circ}$ & $2.79^{\circ}$ \\
	\hline
\end{tabular}
\label{tab:AAVS std dev}
\end{table}

\begin{figure}[htb]
	\centering
	\subfigure[]{
		\includegraphics[width=0.9\linewidth]{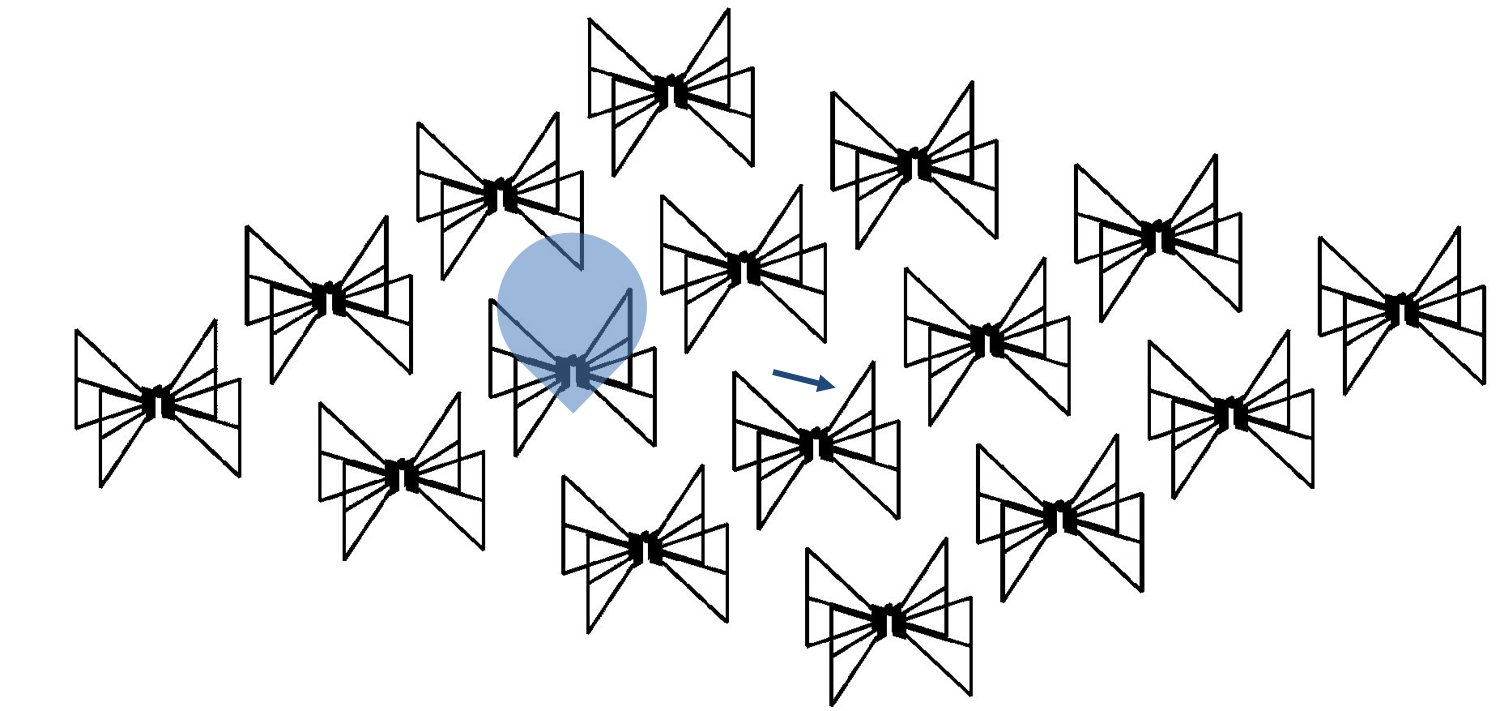}\label{fig:MWA_embed11}} 
	\subfigure[]{
		\includegraphics[width=0.9\linewidth]{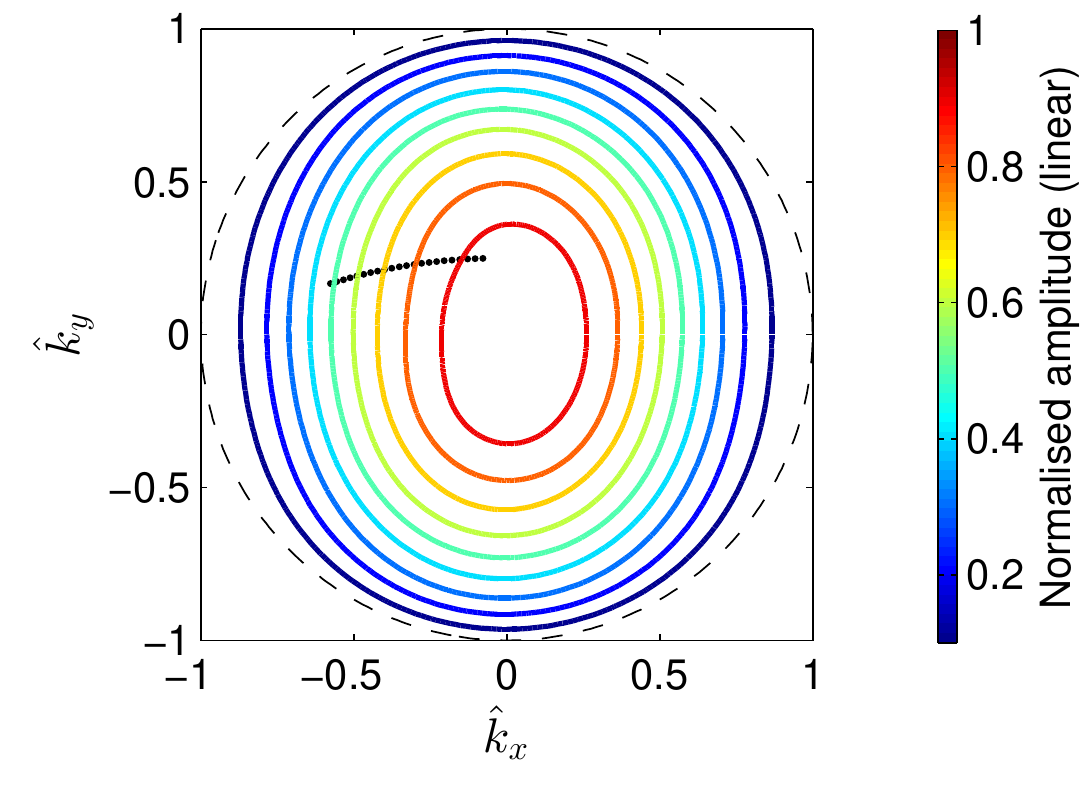}\label{fig:MWA 2D beam}} 
	\caption{Embedded MWA bow-tie element location and beam pattern.~\subref{fig:MWA_embed11} The location of the embedded MWA bow-tie (number 11, blue bubble) in the tile (tile number~041 in MWA numbering scheme). The bow-tie 11 is located at 0.55~m to the east and 0.55~m to the south of the reference center of the array approximately denoted by the tail of the blue arrow. The head of the blue arrow points north.~\subref{fig:MWA 2D beam} The normalized MWA bow-tie embedded element power pattern at 110~MHz. The contour step size is 0.1. Black dotted trace shows the observed HydA trajectory.} 
	\label{fig:MWA beams}
\end{figure}

\begin{figure}[htb]
	\centering
	\subfigure[]{
		\label{fig:MWA cal soln}
		\includegraphics[width=0.9\linewidth]{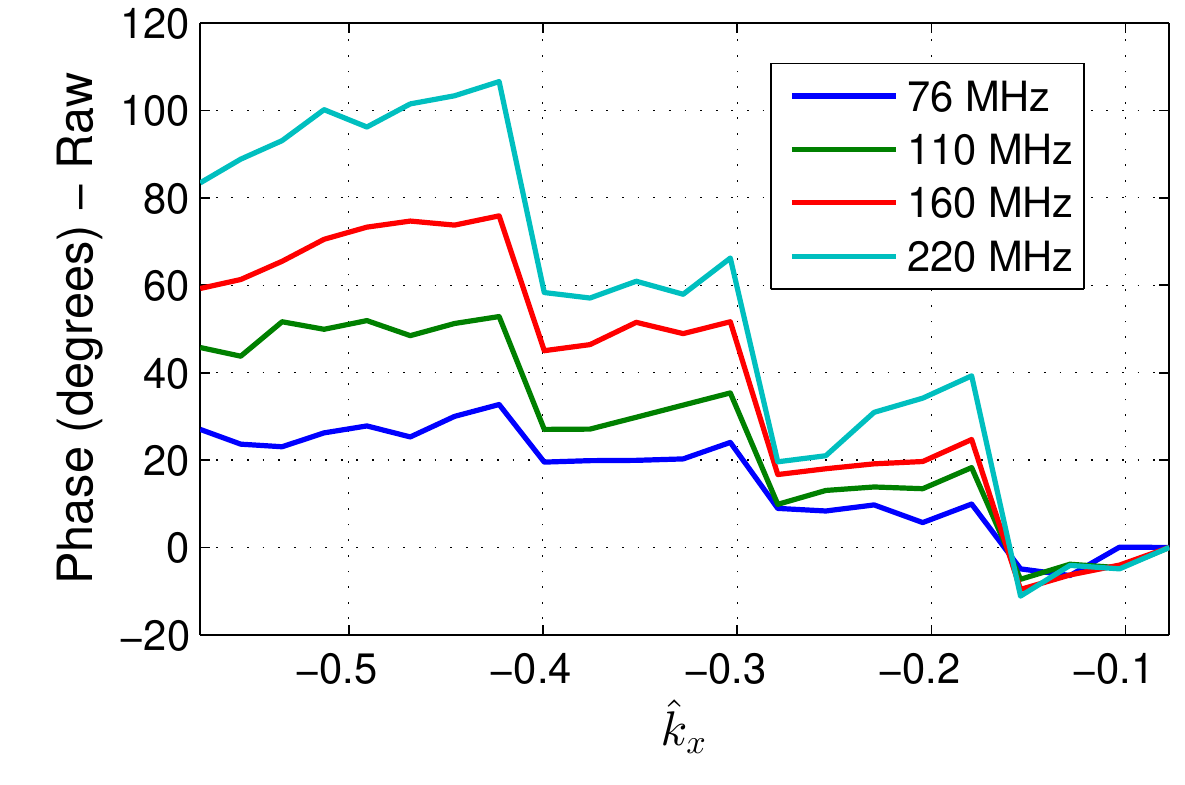}
	} 
	\subfigure[]{
		\label{fig:MWA SMC}
		\includegraphics[ width=0.9\linewidth]{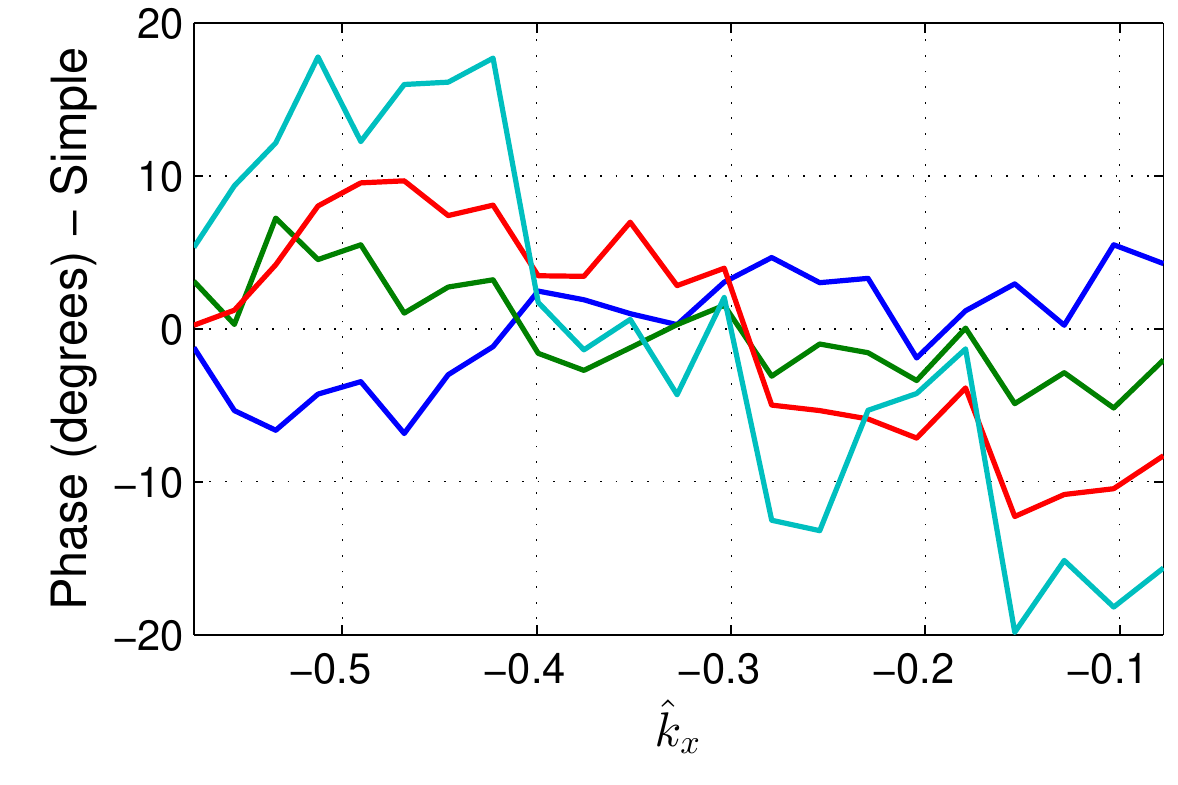}
	}
	
	\subfigure[]{
		\label{fig:MWA GMC}
		\includegraphics[ width=0.9\linewidth ]{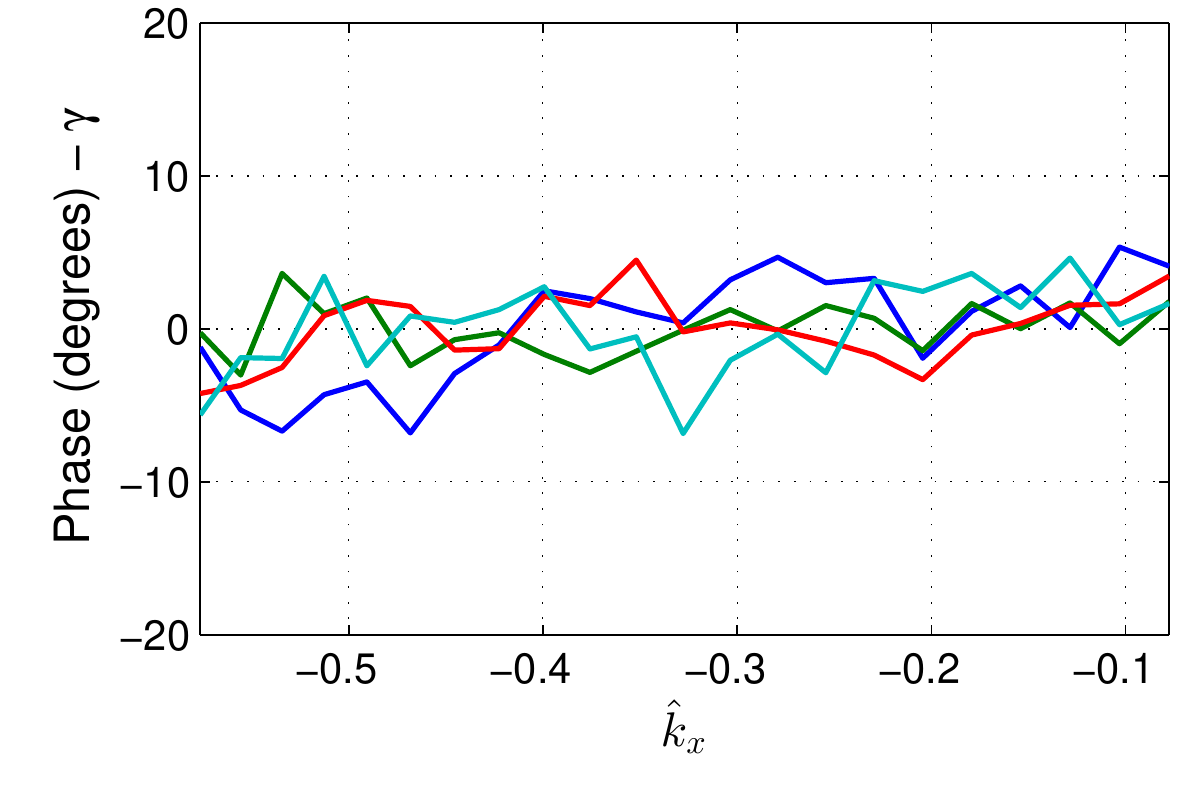}
	}
	
	\caption{Phase of the calibration solutions of MWA bow-tie~11-MWA tile hybrid interferometer with E-W orientation:~\subref{fig:MWA cal soln} raw calibration solution,~\subref{fig:MWA SMC} corrected phase solution using simple geometric model, ~\subref{fig:MWA GMC} corrected using $\gamma$ factor. \label{fig:MWA soln comparison}}
\end{figure}

We repeat this hybrid calibration example using an embedded MWA bow-tie located near the middle of the array (bow-tie~11) as shown in Fig.~\ref{fig:MWA_embed11}. The embedded element pattern at 110~MHz is illustrated in Fig.~\ref{fig:MWA 2D beam}. Fig.~\ref{fig:MWA soln comparison} reports the raw phase calibration solution and corrections with geometric model and $\gamma$ factor. The residual phase of $\sim35^{\circ}$ and phase steps at 220~MHz in Fig.~\ref{fig:MWA SMC} are largely removed in Fig.~\ref{fig:MWA GMC}.

Tab.~\ref{tab:MWA std dev} summarizes the results from Figs.~\ref{fig:MWA SMC} and~\ref{fig:MWA GMC} in terms of standard deviations. We see that $\gamma$ factor correction makes significant improvements at 160~MHz and higher. This is due to the physical size of the MWA bow-tie which is electrically small at $\lesssim$110~MHz~\cite{SutOSu15}, such that mutual coupling effects are not apparent.  This is consistent with Fig.~\ref{fig:MWA 2D beam} which suggests that the MWA bow-tie embedded pattern is less distorted than the embedded SKALA pattern at 110~MHz in Fig.~\ref{fig:AAVS 2D beam}. We conclude from these examples that significant improvement to the calibration phase solution is achievable by applying the $\gamma$ correction, subject to accurate modeling of the array.

\begin{table}[htb]
	\centering
	\caption{MWA bow-tie~11 and MWA tile hybrid calibration: standard deviation of phase residues.}
\begin{tabular}{|c|c|c|}
	\hline
	Frequency & Geometric & $\gamma$ correction \\
	\hline
	75.5 MHz &$3.69^{\circ}$ & $3.67^{\circ}$ \\
	\hline
	110 MHz &$3.33^{\circ}$ & $1.73^{\circ}$ \\
	\hline
	160 MHz &$7.19^{\circ}$ & $2.26^{\circ}$ \\
	\hline
	220 MHz &$12.27^{\circ}$ & $2.95^{\circ}$ \\
	\hline
\end{tabular}
\label{tab:MWA std dev}
\end{table}

\subsection{Polarization Mismatch in Array Sensitivity Measurement}
\label{sec:sens}
Array sensitivity is an important in-situ measurement parameter as it reflects the minimum detectable flux density. Sensitivity is typically expressed as the ratio of antenna aperture area to system temperature ($A_{e}/T_{\mathrm{sys}}$), or as quantity commensurate to its reciprocal, the system equivalent flux density (SEFD). Sensitivity and beam pattern characterization based on sensitivity measurement of AAVS0.5 over the trajectory of HydA has been extensively reported in~\cite{7293140}. Although that work involved hybrid interferometry, we assumed based on the designs of the SKALA and MWA bow-tie (both dual-linearly polarized antennas) that polarization mismatch was insignificant. 

The discussion in Sec.~\ref{sec:AUTcal} suggests that AUT calibration solution is affected by the polarization mismatch. Since SEFD is commensurate to the standard deviation of the calibration solution~\cite{1999ASPC..180..171W, 7293140}, we expect similar effect due to polarization mismatch. We will now quantify the impact of polarization mismatch on sensitivity measurement obtained via AUT calibration solution. The calibration solution is obtained by dividing \eqref{eqn:AAVS_MWA_3} as follows
\begin{eqnarray}
\frac{\langle v_{1}v_{2}^{*}\rangle}{g_{2}^{*}\left\|\mathbf{j}_{2}\right\| \left\| \mathbf{j}_{1}\right\| g_{1}\cos\alpha~e^{j\gamma}}=I/2 
\label{eqn:AAVS_MWA_vis_cal}
\end{eqnarray}
 It can be shown (see Appendix) that the standard deviation of the calibration solution is given by
\begin{eqnarray}
\sigma\big[\Re(\mathcal{V}_{\mathrm{cal}})\big]&=&\sigma\big[\Im(\mathcal{V}_{\mathrm{cal}})\big] \nonumber \\
&=&\sqrt{\frac{2k^{2}}{B t_{\mathrm{acc}}}\frac{T_\mathrm{sys2}}{A_{\mathrm{e}2}}\frac{T_\mathrm{sys1}}{A_{\mathrm{e}1}\cos^{2}\alpha}}
\label{eqn:sens_cal}
\end{eqnarray}
where $\mathcal{V}_{\mathrm{cal}}$ is the left hand side of (\ref{eqn:AAVS_MWA_vis_cal}), $B$ is the signal bandwidth, $t_{\mathrm{acc}}$ is the integration time and $T_{\mathrm{sys}1,2}/A_{\mathrm{e}1,2}$ is the array sensitivity. Hence, the measured array sensitivity is reduced by $\cos^{2}\alpha$ (polarization mismatch) factor as expected. 

At 220 MHz, Fig.~\ref{fig:cos_220} suggests that the sensitivity reduction should be very small ($\cos\alpha>0.95$) for the HydA's trajectory for a single MWA bow-tie and SKALA antennas, as the polarizations are well-matched. We have verified this with full-wave simulation of AAVS0.5 and the MWA tracking HydA along the trajectory reported in~\cite{7293140}. We find well-matched polarization with $\cos\alpha>0.95$ in the main lobe and near sidelobes of AAVS0.5 at 220~MHz~\cite{7303648}. This finding is consistent with our initial assumption. 

\section{Conclusion}
\label{sec:concl}
Hybrid array complex gain calibration of an AUT with $N-1$ identical arrays using a single bright source results in direction-dependent amplitude and phase error factors. This is different from the homogeneous case where only the amplitude factor is direction-dependent. The amplitude error factor is the polarization mismatch factor of the AUT to the identical arrays. This can be minimized by polarization matching of radiated far-fields in the directions of interest. The phase error term is due to movement of antenna/array phase center of the AUT relative to that of the identical arrays. This factor can be characterized and corrected via full-wave EM simulation. We have demonstrated successful phase corrections of a hybrid interferometer involving an AUT and the MWA observing Hydra~A as a calibrator source. We achieve residual phase standard deviations of less than $3^{\circ}$ for an embedded SKALA antenna and an embedded MWA bow-tie antenna at 220~MHz.

\section*{Appendix}
\subsection{\added{Standard Deviation of Calibrated Visibility}}
\label{sec:sens_derive}
\begin{figure}[htb]
	\begin{center}
	\includegraphics[width=3.0in]{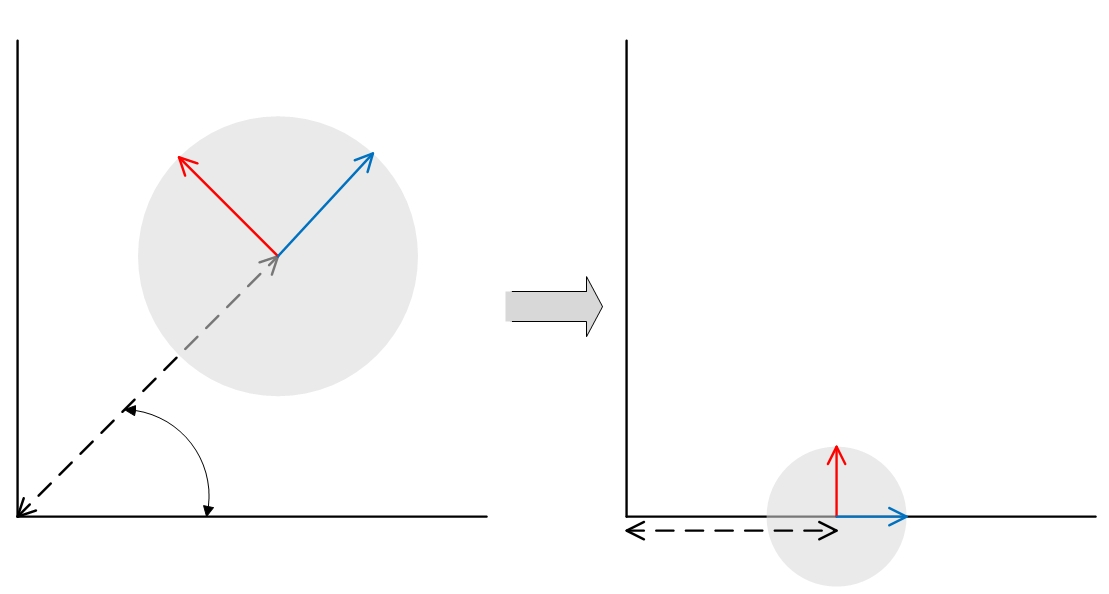}
	\end{center}
\caption{A graphical depiction of the calibration process on the complex plane. The $x-$axis is the real axis and the $y-$axis is the imaginary axis. The figure on the left is the uncalibrated product $ v_{1}v_{2}^{*}= |v_{1}v_{2}^{*}|e^{j\angle v_{1}v_{2}^{*}}$ and the figure on the right is the calibrated product (\ref{eqn:AAVS_MWA_vis_cal}).}
\label{fig:complex_gain_Cal}
\end{figure}
The calibration process in nicely described with the aid of the graphical representation in Fig.~\ref{fig:complex_gain_Cal}. One can think of the process as a two-step rotate then scale operation. The rotate operation removes the $\angle \left\langle v_{1}v_{2}^{*}\right\rangle$ (the angle indicated on the left graph in Fig.~\ref{fig:complex_gain_Cal}) such that the data straddles the real axis. Assuming that the correlated signal is much fainter than the system noise~\cite{1999ASPC..180..171W, 7293140} \added{(which is indeed the case for HydA observation with the MWA~\cite{7293140})}, the noise variance of the real and imaginary components are given by:
\begin{eqnarray}
\sigma^{2}\big[\Re(\mathcal{V}_{\mathrm{uncal}})\big]&=&\sigma^{2}\big[\Im(\mathcal{V}_{\mathrm{uncal}})\big] \nonumber \\
&=&k^{2}\left|g_{2}\right|^{2}T_\mathrm{sys2}\left|g_{1}\right|^{2}T_\mathrm{sys1}
\label{eqn:vac_uncal}
\end{eqnarray}

Next, the scaling operation divides the rotated product with the magnitude of the denominator of (\ref{eqn:AAVS_MWA_vis_cal}). The variance of the rotated and scaled product is
\begin{eqnarray}
\frac{\sigma^{2}\big[\Re(\mathcal{V}_{\mathrm{uncal}})\big]}{(\left|g_{2}\right|\left\|\mathbf{j}_{2}\right\| \left\| \mathbf{j}_{1}\right\| \left|g_{1}\right|\cos\alpha)^{2}}=k^{2}\frac{T_\mathrm{sys2}T_\mathrm{sys1}}{\left\|\mathbf{j}_{2}\right\|^{2} \left\| \mathbf{j}_{1}\right\|^{2} \cos^{2}\alpha}
\
\label{eqn:sigma_cal}
\end{eqnarray}
After accounting for averaging, it can be shown with some algebra\cite{1999ASPC..180..171W} that  equation (\ref{eqn:sens_cal}) is obtained.

\subsection{\added{Investigation into the Residual Phase of Embedded SKALA11 at 160~MHz}}
\begin{figure}[htb]
	\subfigure[]{
		\includegraphics[width=0.9\linewidth]{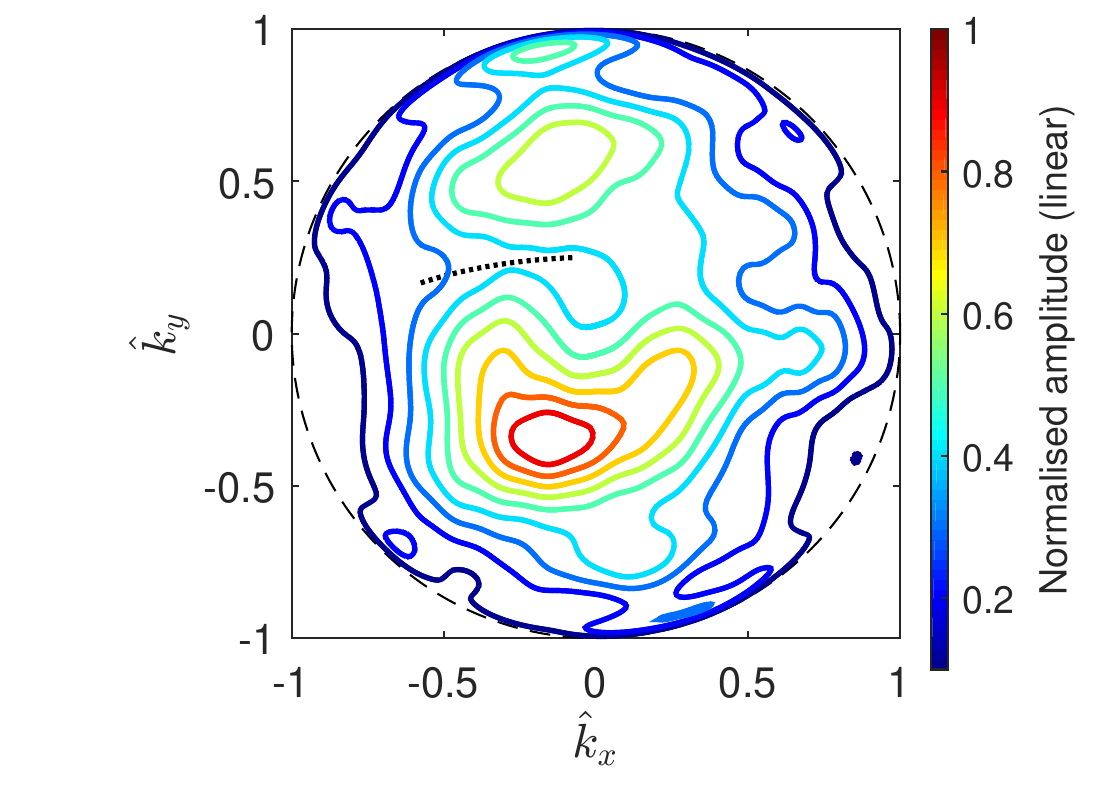}\label{fig:SKALA_160MHz}} 
	\subfigure[]{
		\includegraphics[width=0.9\linewidth]{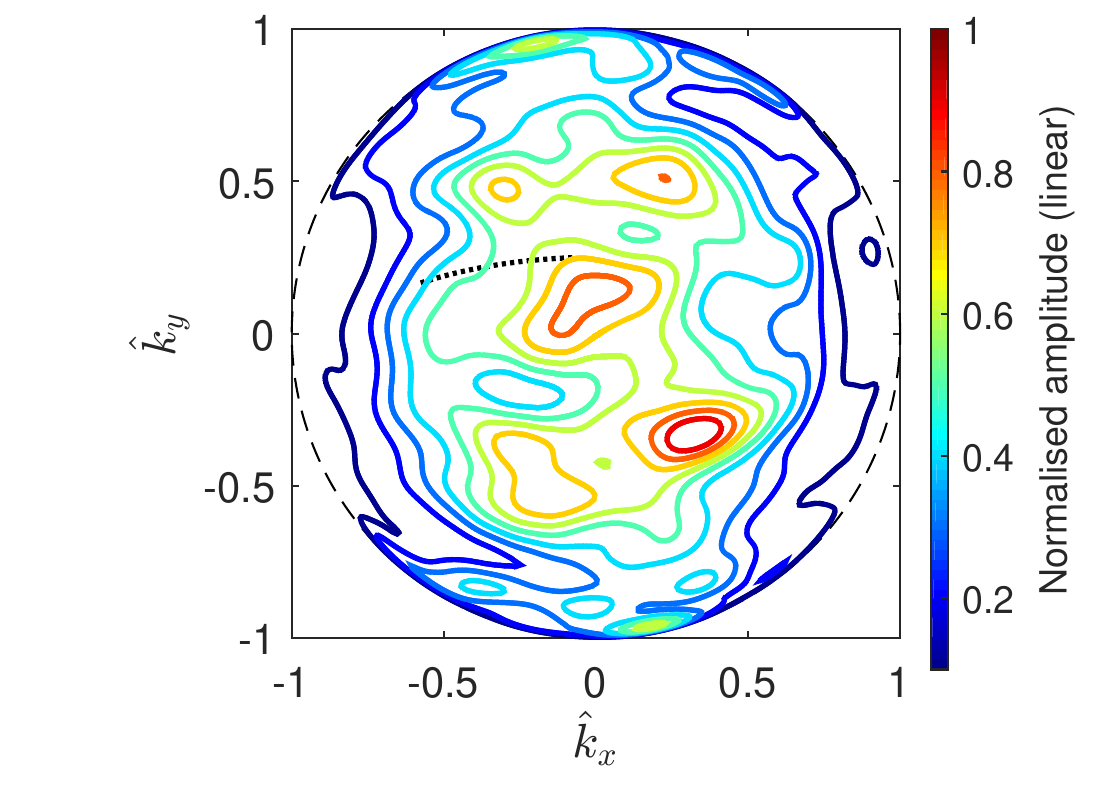}\label{fig:SKALA_220MHz}} 
	\caption{\added{Contour plot of normalized power pattern of SKALA~11 at 160 MHz \subref{fig:SKALA_160MHz} and 220 MHz \subref{fig:SKALA_220MHz}. The dotted lines shows the track of HydA during the observational period. The contour step is 0.1.}} 
	\label{fig:SKALA_beams}
\end{figure}

\added{Field inspection of SKALA11 has ruled out mechanical issues such as damage or visible misalignment. Next, we turn our attention to the embedded element pattern of SKALA~11. Fig.~\ref{fig:SKALA_beams} reports the trajectory of HydA superimposed on the normalized embedded power pattern. At 160~MHz, HydA is tracking along a ``valley'' (with contour values of 0.3 to 0.4) between two local peaks. At 220~MHz or 110~MHz (see Fig.~\ref{fig:AAVS 2D beam}) the trajectory steadily climbs toward a peak (or the vicinity thereof). The same steady climb is also observed with HydA in the MWA bow-tie~11 power pattern at 160~MHz and 220~MHz (not shown). These results suggest that the EM model is likely less accurate in the ``valley'' region than in the peak region (similar to higher sensitivity to tolerance in the vicinity of nulls of a phased array response).} 

\section*{\added{Acknowledgment}}
\added{This scientific work makes use of the Murchison Radio-astronomy Observatory, operated by CSIRO. We acknowledge the Wajarri Yamatji people as the traditional owners of the Observatory site. Support for the operation of the MWA is provided by the Australian Government (NCRIS), under a contract to Curtin University administered by Astronomy Australia Limited. We acknowledge the Pawsey Supercomputing Centre which is supported by the Western Australian and Australian Governments.}


\end{document}